# Strengthening e-Education in India using Machine Learning

Naheed Khan
K J Somaiya Institute of Engineering & Information Technology, Sion, Mumbai
University of Mumbai, India
naheed.khan@somaiya.edu

Darshan Bhanushali
K J Somaiya Institute of Engineering & Information Technology, Sion, Mumbai
University of Mumbai, India
darshan.jb@somaiya.edu

Shreya Patel
K J Somaiya Institute of Engineering & Information Technology, Sion, Mumbai
University of Mumbai, India
shreya.bp@somaiya.edu

Radhika Kotecha
K J Somaiya Institute of Engineering & Information Technology, Sion, Mumbai
University of Mumbai, India
Radhika.kotecha@somaiya.edu

*Abstract-* *e-Education has developed as one of the most encouraging territories. The Indian Government is investing all amounts of energy to improve education among the residents of the nation. School and graduate understudies are focused on, however the stage is being created for all the residents seeking to learn. Without a doubt, the objective is to build the quantity of literates with advanced education. To accomplish the equivalent, propels in Data and Correspondence innovation are being utilized in the education division, which has cleared route for e-Training in India as well. To help educators in concentrating more on more current viewpoints, their excess work can be disposed of utilizing Machine Learning (ML). Difference to programming, ML deals with information and answers to create rules. In the event that Machine Learning is tackled effectively, it can setup the training division and contribute essentially to the development of the country. Hence, the work presented in this paper fortifiews e-Education in India utilizing Machine Learning. For the most part, three concerns are focused to be tended to: Personalized recommendation of course and Customized teaching methodology. The work proposes utilizing developmental methodology of hereditary calculations for improving conventional procedures. Implementation and experiments presented in the paper verify the viability of proposed calculations.*

*Keywords-* **e-Education, Data mining, e-learning, Regression, Machine learning**

## I. INTRODUCTION

There are several sectors where Machine Learning [1] is playing an important role. Various banks are adopting Machine Learning algorithm for fraud detection and prevention, financial organizations use Machine Learning for an observation of customers behavior, healthcare sector uses different Machine Learning-based models are used for the purpose of early prediction of disease and to find their cure, manufacturing industries are adopting automation based on Machine Learning, retail industries use various Machine Learning technologies to capture, analyze and use data to personalize the shopping experience of their customers, etc. [2].

One of the major areas where technology can focus is e-education or e-learning. It is the delivery of education or any type of training by electronic teaching methods. This electronic method can be a computer or a smartphone where teaching material is accessed by the use of the internet [3].

Currently, there are several issues ranging from quaint courses, lack of hands-on practical experience, lack of quality educators, then forth. This points out the sad and scary state of India's education system and presents the difficulty graduates face while looking for a job. e-Education is gaining popularity among people. e-Education is helping people to learn and get knowledge as per their choice and need. e-Education also provides the updated technology course that is currently in demand and used in the world. It also helps a person to learn with his speed and time [4]. By using on-line learning facilities, students have access to good quality education courses, imparted by seasoned and experienced professors and professionals, on their fingertips. While online learning is bringing quality educational courses to students, it is also making the entire learning process 'fun'.e-learning resources, students will access it at a far lower price through e-learning

In the world of technology e-Education plays a very important role in educating different types of people that have different skill sets and learning speed.The problem with today's e-Education is a particular person starts an online course that may be suggested by his/her friends or teachers. In this the person may not be able to catch up with that course and its way of teaching .This problem is also known as "One size fits all"Concept" Hence it is essential to first check that a person should have basic knowledge of that course.". People often start an online course and end up quitting in the middle. This is because the person can not catch up with the method of teaching or the person does not have enough knowledge or level for that particular course.

We have used Simple decision tree classifier that uses the supervised classification on the static dataset, for the purpose of course recommendation as it has given us quite good accuracy as compare to the remaining two algorithms i.e. Random Forest and Naive bayes classifier.[5,6]

We have done a proper study of all the three algorithms to get the insight about each and every aspect of it. We chose the Decision tree on the basis of its accuracy as well as it can handle high dimensional data,it can also handle continuous and categorical data more efficiently which is a difficult task in remaining two algorithms. We got almost the same results by SDT and Random Forest but the accuracy of the decision tree was 75% which is quite higher than the others.

*A. Contribution of this Work*

1. We are applying different data mining algorithms on the data of the person and suggest which course is appropriate for him based on his background knowledge. We will also be able to provide online teaching based on his learning speed and the methodology of teaching he prefers. This will significantly improve the e-Education system of learning.

2. Through e-learning, students will target turning into 'employable' and build their dream career.. This is simply the start of the revolutionary conception of e-learning that has the potential to disrupt the education system in India and provide a better way in learning environment for the students.

This paper is organized as follows: In Section 1, we provide a brief description of the basic concepts used in the paper. Section 2, summarizes the existing work related to e-Education in India and Machine Learning. The proposed algorithm for strengthening e-Education in India using Machine Learning is stated in Section 3. Section 4, states the experimental setup that demonstrates the performance. Section 5, the result of the entire proposed system is demonstrated. Section 6, concludes the work and gives highlights for future work.

## II. RELATED WORK

This section presents, in brief, existing literature related to the use of Machine Learning in e-education.

Researchers R. Banu, R.Ravanan [7] have given a brief knowledge about e-learning and how it can be used to empower the rural students, the authors have also given the information about how data mining algorithm will be helpful to e-learning and how it can be more useful in e-learning, it also gives us knowledge about KDD that is Knowledge Discovery and Data Mining and how it can be used to preprocess, sample and project the data prior to the data mining steps.

In [8], S.Senthil, W. Lin, have implemented and compared various data mining algorithms such as Bayes-based, ANN-based, Regression-based, SVM-based, Instance-based, Tree-based and Rule-based classification algorithms on the student data, the algorithms were compared in terms of accuracy rate, precision rate, AUC and model building time.As per author RandomForest is the best algorithm in terms of accuracy, precision, and AUC in predicting students' academic results on their datasets.

N. Anh [9], proposed a robust approach for this inevitable challenging problem. Their work gives the semantics of incomplete data in the educational sector on the application part and the two-phase characteristics of the classification task on the technical side. As a result of this study on real educational data sets with different parts of incomplete data, they found that the different approaches for incomplete data handling.

In [10],A. Ahadi et al., introduced an approach that combines whether or not a student-produced a correct solution to an online exercise with information on the number of attempts at the exercise submitted by the student. They used the data collected from students in an introductory Java course to assess the value of this approach.They found that if the number of attempts is correlated than student performance is better in the final exams.

In [11] C. Romero, have introduced EDM and describes the different groups of users, types of educational environments, and the data they provide and then gives insight how data mining techniques have solved educational issues.

M. Santillan et al. [12], have tried to build more accurate classification models in order to predict the output, analyzing the interaction in an incremental way and compare the total interaction models with incremental interaction models.

In [13] the authors have elicited knowledge from the information collected from faculty and students of two different educational institutes. Also they have defined Multilevel Hierarchy of training quality.

A. Pal, S. Pal [14], have described the use of data mining techniques t.o improve the efficiency of academic performance in educational institutions. The basic techniques used in this paper are ID3, c4.5 and Bagging.

In [15] B.Namratha, N. Sharma have discussed educational data mining, its broad application areas, benefits of educational data mining, challenges and barriers to successful application of educational data mining and the new practices that have to be adopted in order to successfully employ educational data mining and learning analytics for improving teaching and learning.
Based on the above mentioned studies we have proposed a approach to develop a web portal.

### III. PROPOSED APPROACH
`
Although shifting from traditional education to e-education is a much needed aspect in India, to the best of our knowledge there is no such real time system that analyzes e-education. Basically e-learning is a new way of providing quality of knowledge and education to people to interact with web based portals. It is that teachers make simple statistical checking and analysis on the results of the students at the end of a course, and this is useful for the evaluation of that course. However a more powerful use of statistics is used to improve the methodology of learning.[16]

Currently there is an e-education system which will give you content like videos, online lectures etc. All these systems work either with databases or log files and then use this data to provide different resources online. There exist many systems like youtube, udemy etc., which give online videos for study. During the last few years researchers have carried out studies to develop some models to improve these systems which can give support to teachers as well as students but these systems are not actually implemented in real time.

The proposed approach is a web portal where a user can register and login. Users need to give a preliminary test based on which the system will evaluate his/her score in order to suggest him/her different courses.The website will make use of the model that we will build using the Machine Learning algorithms to provide the output.

As shown in Figure 1, initially on this web portal there will be a prerequisite test related to different subjects like DSA, Java, ML, etc. After getting answers from students, they will come to know about his/her area of interest and proficient area. Based on all the information which we have collected earlier by giving tests, the system will be able to suggest students some courses related to his/her area of interest.

In the second phase the portal shows them videos related to their respective course to know their learning pace whether they understand theory easily or they need some examples or diagrams etc.After showing this, there will be some series of questions to check their performance score in that particular course.

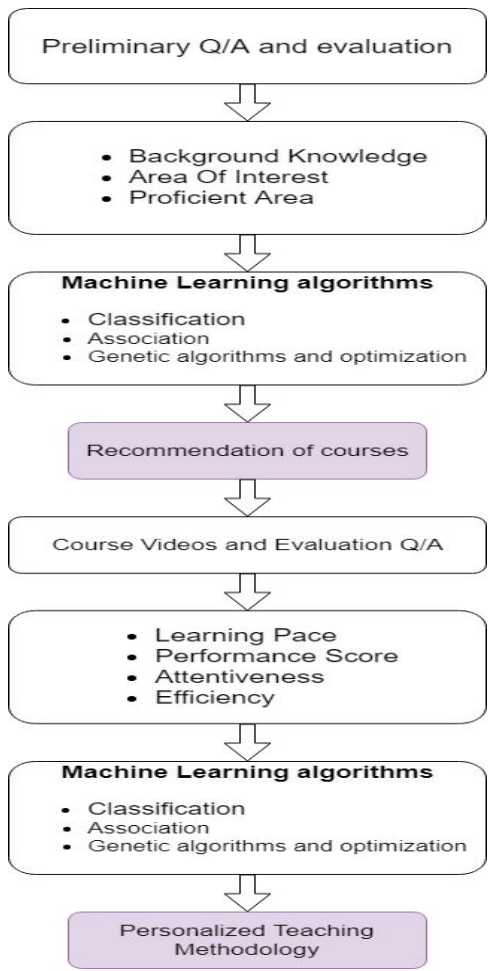

Figure 1: Proposed System

IV. EXPERIMENTAL SETUP

To demonstrate the effectiveness of the proposed algorithm, we performed empirical studies with different data sets. In this section, we present the dataset used, implementation details and report evaluation results.

*A. Dataset*

Considering the need to address the web application described in the proposed approach of this paper, we considered several different attributes with a class label Course Recommendation along with the level of course. We have taken approximately five thousand instances..
We have considered the subject, Basic level correct answers (BLA), Medium level correct answer (MLA), High level correct answer (HLA) and average score of students as a main attributes. Based on the BLA, MLA and HLA we have calculated the average score.

*B. Baseline methods for comparison*

To demonstrate the effectiveness of the proposed algorithm, we compare its performance with three different methods. Details of this method are described in the following section.

**Simple Decision Tree -** Decision tree algorithm is a supervised learning algorithm. It uses tree representation to solve the problem. In SDT attributes are represented by internal nodes whereas leaf nodes represent the class label.

**Random Forest -** Random forest is a combination of several simple decision trees. Each individual tree in the random forest splits out a class prediction the class with the most votes becomes the model prediction.

**Naive Bayes -** Naive Bayes is a simple classification algorithm which is used for predictive modeling. Naive Bayes algorithm gives us a way by which we can calculate the probability of hypothesis based on prior knowledge.

*C. Software Details*

In this section we have described different softwares and the programming languages we have used in order to develop the web portal.

**XAMPP -** XAMPP stands for Cross-Platform, Apache MariaDB PHP Perl. It is a lightweight Apache distribution that makes it very easy for developers to create a local web server for testing and deployment purposes.

**Laravel -** Laravel is an open source php framework which works on a Model View Controller design pattern. It helps in creating a rich web application by utilizing the existing components of different framework.

**Python -** Python is a programming language which is used on a server to create web applications. Python has a very simple syntax and can be used on different platforms. It can be used in a procedural way, Object oriented way or a functional way.

**PHP -** PHP stands for Hypertext Preprocessor. It is a scripting language. These scripts are executed on the server. PHP is used to generate dynamic page content.

**Weka -** Weka is an open source tool which is used for data preprocessing, data visualization and implementation of several Machine Learning algorithms. It is basically used to solve different problems based on data mining or Machine Learning.

## V. RESULTS

In this section we demonstrate the results observed through different algorithms. We have used a hold-out method of evaluation where 80% of the data is used stream for training the classifier and the rest for testing.

The Accuracy and F-measure [13] are used for evaluating performances of algorithms

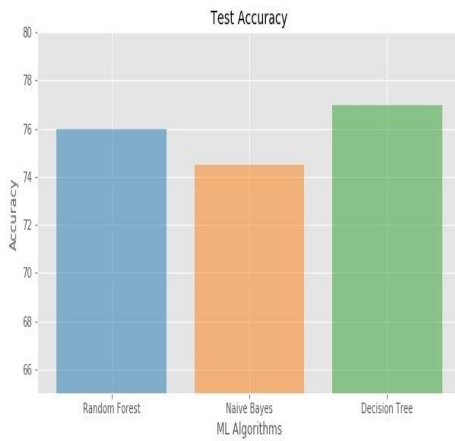

Figure 2: Test Accuracy

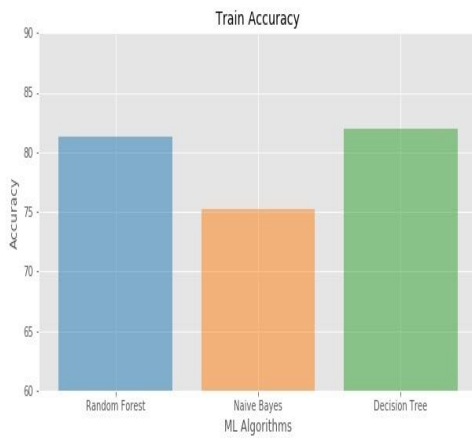

Figure 3: Training Accuracy

Figure 2 & Figure 3 represents the accuracy results of evaluation as testset and training set respectively.

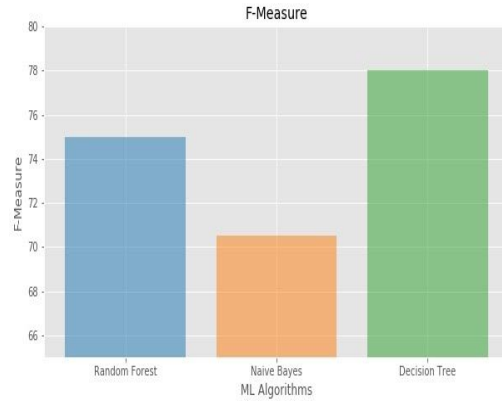

Figure 4: F-Measure

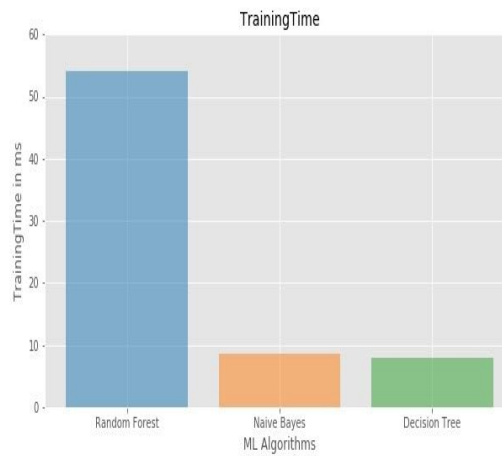

Figure 5: Training Time

Figure 4 & Figure 5 demonstrate the results of F-Measure and Training Time.

For F-Measure,

$$\text{F-Measure} = \frac{(2 * \text{Precision} * \text{Recall})}{(\text{Precision} + \text{Recall})} \quad (1)$$

For Accuracy,

$$\text{Accuracy} = \frac{(TP + TN)}{(TP + TN + FP + FN)} \quad (2)$$

where TP, FN, FP and TN represent the number of true positives, false negatives, false positives and true negatives, respectively.

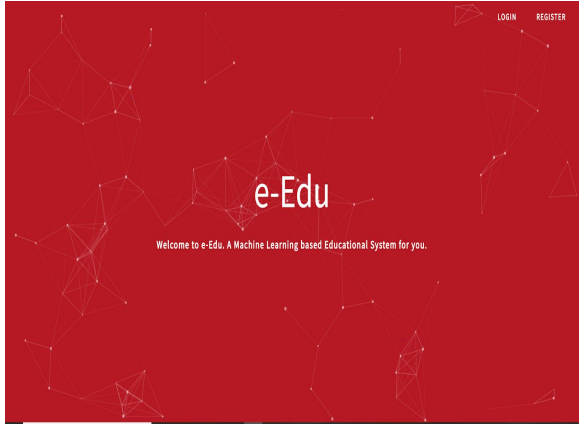

Figure 6: Home Page

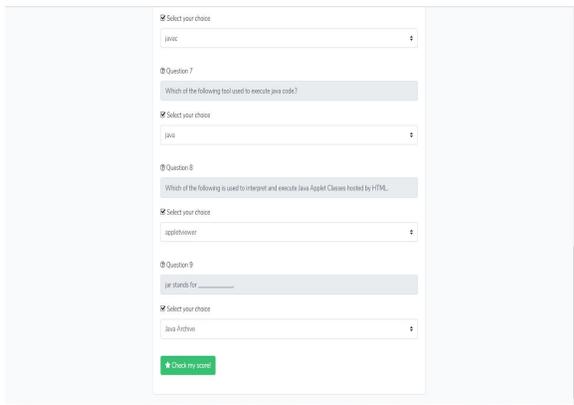

Figure 7: Q&A Evaluation Page

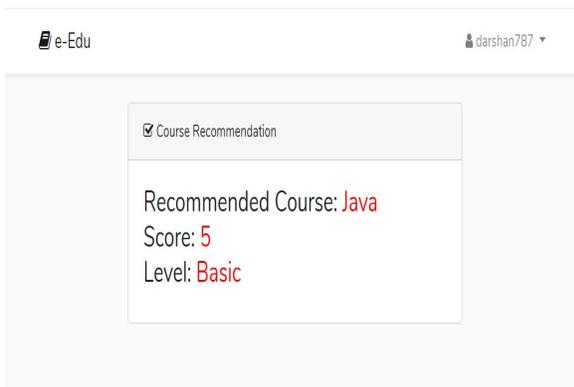

Figure 8: Course Recommendation with Level

Figure 6, 7 & 8 shows the actual implementation of web application. Course recommendation and level of course is based on the score of a student.

## VI. CONCLUSION & FUTURE WORK

At present, educational data mining is paid great attention by most of the researchers, since it evolves in improving the current educational systems. This report concludes that Data Mining strongly contributes to the betterment of imparting quality in higher education by screening several applications in new dimensions. This work is an effort to enhance the traditional educational process via strategic roadmap of data mining functionalities.The performance, success of students in the examination.

There are several things which will be implemented in future such as course videos based on evaluation of present knowledge of students thereby inculcating additional knowledge and providing them a scope to further expand their arsenal of knowledge.